\documentclass{elsart}

\usepackage{amsmath,amssymb}

\allowdisplaybreaks

\journal{Physics Letters A}
\newcommand{\cN}{\mathcal{N}}
\newcommand{\cV}{\mathcal{V}}
\newcommand{\cW}{\mathcal{W}}
\newcommand{\bH}{\mathbf{H}}
\newcommand{\bQ}{\mathbf{Q}}
\newcommand{\bbR}{\mathbb{R}}
\newcommand{\bbC}{\mathbb{C}}
\newcommand{\ee}{\mathrm{e}}
\newcommand{\dd}{\mathrm{d}}

\newcommand{\rnu}{\sqrt{\nu}}

\begin{document}

\begin{frontmatter}

\title{Simultaneous Type A $\cN$-fold Supersymmetry with Two Different
 Values of $\cN$}
\author{Choon-Lin Ho},
\author{Toshiaki Tanaka}
\ead{ttanaka@mail.tku.edu.tw}
\address{Department of Physics, Tamkang University,
 Tamsui 25137, Taiwan, R.O.C.}

\begin{abstract}

We investigate one-dimensional quantum mechanical systems which
have type A $\cN$-fold supersymmetry with two different values
of $\cN$ simultaneously. We find that there are essentially four
inequivalent models possessing the property, one is conformal,
two of them are hyperbolic (trigonometric) including Rosen--Morse
type, and the other is elliptic.

\end{abstract}


\end{frontmatter}

\section{Introduction}
\label{sec:intro}

Discovery of quasi-exact solvability in quantum mechanics
\cite{TU87,Us94} has promoted the investigation of quantum systems
which admit exact solutions. Recently, this concept was combined
within the framework of $\cN$-fold supersymmetry \cite{AST01b},
which is a natural generalization \cite{AIS93} of ordinary
supersymmetric quantum mechanics \cite{Wi81}. Furthermore,
a systematic algorithm for constructing an $\cN$-fold supersymmetric
system was established based on the connection between (weak)
quasi-solvability and $\cN$-fold supersymmetry \cite{GT05}.
Up to now, three different families of $\cN$-fold supersymmetric
systems have been found for arbitrary finite integer $\cN$,
namely, type A \cite{AST01a,Ta03a}, type B \cite{GT04}, and
type C \cite{GT05}, which have correspondence with the
classification of second-order linear differential operators
preserving a monomial-type vector space \cite{PT95}.

One of the intriguing aspects of $\cN$-fold supersymmetry is its
dynamical breaking. We can actually observe the phenomenon through
the nonperturbative effect due to the quantum tunneling
\cite{AKOSW99,ST02}. In particular, there are several characteristic
features which are different from the case of ordinary supersymmetric
quantum mechanical systems \cite{AST01b}. In the case of $\cN$-fold
supersymmetry, semi-positive definiteness of the spectrum is not
guaranteed, more than one state can be invariant with respect to one
supercharge, and so on. Actually, it is shown that the latter fact
can lead to a novel phenomenon, namely, partial breaking of $\cN$-fold 
supersymmetry \cite{GT05}. Hence, it is interesting to find out
a realistic physical system with $\cN$-fold supersymmetry.\footnote{
The models in Ref.~\cite{KP01b} have unphysical magnetic moment.}

On the other hand, ordinary supersymmetric quantum mechanical systems
have been extensively studied more than two decades (for a review,
see e.g. \cite{CKS95}). In particular, several realistic physical
systems having ordinary supersymmetry have been found not only for
nonrelativistic scalar systems but also for Pauli and Dirac systems
\cite{AC79,Ui84,GZ88,AI88,CKMW88,Se90,BR99,HR03,HR04}. Hence, it
suggests that we shall begin with looking for an ordinary supersymmetric
system which has $\cN$-fold supersymmetry as well, as a starting
point of the aforementioned purpose. This observation naturally leads
us further to consider more general situation where a system has
simultaneous $\cN$-fold supersymmetry with two different values of
$\cN$. In this article, we show a family of that kind of quantum
mechanical systems.

In the following, we first review the definition of $\cN$-fold
supersymmetry and extend it to the concept of simultaneous $\cN$-fold
supersymmetry with two different values of $\cN$. In Section
\ref{sec:typeA}, we fully investigate the condition for simultaneous
$\cN$-fold supersymmetry with respect to two different type A
$\cN$-fold supercharges and classify the models possessing that
property. Finally, we summarize the results and discuss some future
problems.

\section{$\cN$-fold Supersymmetry and Quasi-solvability}
\label{sec:Nfold}

First of all, we shall review the concept of $\cN$-fold supersymmetry
in one-dimensional quantum mechanics. Let $q$ denote a bosonic
coordinate, and let $\psi$ and $\psi^{\dagger}$ be fermionic
coordinates satisfying
\begin{align}
\{\psi,\psi\}=\{\psi^{\dagger},\psi^{\dagger}\}=0,
 \qquad\{\psi,\psi^{\dagger}\}=1.
\end{align}
We define a super-Hamiltonian $\bH$ by
\begin{align}
\bH=H^{-}\psi\psi^{\dagger}+H^{+}\psi^{\dagger}\psi,
\end{align}
where $H^{\pm}$ is a pair of ordinary scalar Hamiltonians:
\begin{align}
H^{\pm}=-\frac{1}{2}\frac{\dd^{2}}{\dd q^{2}}+V_{\cN}^{\pm}(q).
\end{align}
With a monic $\cN$th-order linear differential operator
\begin{align}
P_{\cN}=\frac{\dd^{\cN}}{\dd q^{\cN}}+\sum_{k=0}^{\cN-1}w_{k}(q)
 \frac{\dd^{k}}{\dd q^{k}},
\end{align}
$\cN$-fold supercharges $\bQ_{\cN}^{\pm}$ are introduced by
\begin{align}
\label{eq:defbQ}
\bQ_{\cN}^{-}=P_{\cN}^{-}\psi^{\dagger},
 \qquad\bQ_{\cN}^{+}=P_{\cN}^{+}\psi,
\end{align}
where the operators $P_{\cN}^{\pm}$ are defined by
\begin{align}
P_{\cN}^{-}=P_{\cN},\qquad P_{\cN}^{+}=P_{\cN}^{t},
\end{align}
the superscript $t$ denoting the formal transposition. Then,
the system $\bH$ is said to have \emph{$\cN$-fold supersymmetry}
if it satisfies
\begin{align}
\bigl[\bQ_{\cN}^{\pm},\bH\bigr]=0.
\end{align}
One of the most important aspects of $\cN$-fold supersymmetry is that
the component Hamiltonians $H^{-}$ and $H^{+}$ are always \emph{weakly
quasi-solvable} with respect to the operators $P_{\cN}^{-}$ and
$P_{\cN}^{+}$, respectively \cite{AST01b,Ta03a}. That is, $H^{\pm}$
leave the kernels of $P_{\cN}^{\pm}$ invariant:
\begin{align}
H^{\pm}\cV_{\cN}^{\pm}\subset\cV_{\cN}^{\pm},
 \qquad\cV_{\cN}^{\pm}=\ker P_{\cN}^{\pm}.
\end{align}
As a consequence, we can in principle diagonalize algebraically
the Hamiltonians $H^{\pm}$ in the finite $\cN$-dimensional vector
spaces $\cV_{\cN}^{\pm}$, which are thus called
\emph{solvable sectors} of $H^{\pm}$. If the space $\cV_{\cN}^{+(-)}$
is a subspace of a Hilbert space $L^{2}$ on which the Hamiltonian
$H^{+(-)}$ is defined, the elements of $\cV_{\cN}^{+(-)}$ provide
a part of the exact eigenfunctions and thus $H^{+(-)}$ is called
\emph{quasi-exactly solvable}.

Next, the system $\bH$ is said to have \emph{$(\cN_{1},\cN_{2})$-fold
supersymmetry} if it commutes with two different $\cN_{i}$-fold
supercharges ($i=1,2$) \emph{simultaneously}, namely,
\begin{align}
\bigl[\bQ_{\cN_{1}}^{(1)\pm},\bH\bigr]
 =\bigl[\bQ_{\cN_{2}}^{(2)\pm},\bH\bigr]=0.
\end{align}
Without loss of generality, we can assume $\cN_{1}\geq\cN_{2}$.
In this case, it is evident that the components $H^{\pm}$ of the
system $\bH$ preserve two vector spaces $\cV_{\cN_{1}}^{(1)\pm}=
\ker P_{\cN_{1}}^{(1)\pm}$ and $\cV_{\cN_{2}}^{(2)\pm}=\ker
P_{\cN_{2}}^{(2)\pm}$ separately, where $P_{\cN_{i}}^{(i)\pm}$
are components of $\bQ_{\cN_{i}}^{(i)\pm}$ ($i=1,2$), cf.
Eq.~\eqref{eq:defbQ}. Hence, the solvable sectors
$\cV_{\cN_{1}\!,\,\cN_{2}}^{\pm}$ of $(\cN_{1},\cN_{2})$-fold
supersymmetric Hamiltonians $H^{\pm}$ are generally given by
\begin{align}
\cV_{\cN_{1}\!,\,\cN_{2}}^{\pm}=\cV_{\cN_{1}}^{(1)\pm}
 \cup\cV_{\cN_{2}}^{(2)\pm}.
\end{align}

\section{Type A $(\cN_{1},\cN_{2})$-fold Supersymmetry}
\label{sec:typeA}

In what follows, we shall investigate the $(\cN_{1},\cN_{2})$-fold
supersymmetric systems with respect to two type A $\cN_{i}$-fold
supercharges ($i=1,2$), which we hereafter call \emph{type A
$(\cN_{1},\cN_{2})$-fold supersymmetric}. It was briefly discussed
in Ref.~\cite{Ta03a} in a different context, but here we will
make a precise treatment. The component of the type A $\cN_{i}$-fold
supercharge is defined by
\begin{multline}
P_{\cN_{i}}^{(A)}=\biggl(\frac{\dd}{\dd q}+W_{i}(q)
 -\frac{\cN_{i}-1}{2}E_{i}(q)\biggr)\biggl(\frac{\dd}{\dd q}+W_{i}(q)
 -\frac{\cN_{i}-3}{2}E_{i}(q)\biggr)\times\cdots\\
\cdots\times\biggl(\frac{\dd}{\dd q}+W_{i}(q)+\frac{\cN_{i}-3}{2}
 E_{i}(q)\biggr)\biggl(\frac{\dd}{\dd q}+W_{i}(q)
 +\frac{\cN_{i}-1}{2}E_{i}(q)\biggr).
\end{multline}
According to Ref.~\cite{Ta03a}, the necessary and sufficient
condition for type A $\cN_{i}$-fold supersymmetry is the following:
\begin{align}
\label{eq:Apots}
V_{\cN_{i}}^{\pm}(q)=\frac{1}{2}W_{i}(q)^{2}-\frac{\cN_{i}^{\,2}-1}{24}
 \bigl(2\dot{E}_{i}(q)-E_{i}(q)^{2}\bigr)\pm\frac{\cN_{i}}{2}
 \dot{W}_{i}(q)-R_{i},
\end{align}
where the dot denotes derivative with respect to $q$, $R_{i}$ is
a constant, and the functions $W_{i}(q)$ and $E_{i}(q)$ satisfy
\begin{gather}
\label{eq:Acon1}
\biggl(\frac{\dd}{\dd q}-E_{i}(q)\biggr)\frac{\dd}{\dd q}
 \biggl(\frac{\dd}{\dd q}+E_{i}(q)\biggr)W_{i}(q)=0\quad
 \text{for}\quad\cN\ge 2,\\
\biggl(\frac{\dd}{\dd q}-2E_{i}(q)\biggr)
 \biggl(\frac{\dd}{\dd q}-E_{i}(q)\biggr)\frac{\dd}{\dd q}
 \biggl(\frac{\dd}{\dd q}+E_{i}(q)\biggr)E_{i}(q)=0\quad
 \text{for}\quad\cN\ge 3.
\end{gather}
In order that the pair of potentials $V_{\cN_{i}}^{\pm}$ be type A
$(\cN_{1},\cN_{2})$-fold supersymmetric, $V_{\cN_{1}}^{\pm}$ and
$V_{\cN_{2}}^{\pm}$ must be identical up to an additive constant,
namely, $V_{\cN_{1}}^{\pm}+R_{3}=V_{\cN_{2}}^{\pm}\equiv
V_{\cN_{1}\!,\,\cN_{2}}^{\pm}$, where $R_{3}$ is a constant. From
Eq.~\eqref{eq:Apots} we immediately have
\begin{align}
\label{eq:cond1}
\cN_{1}\dot{W}_{1}&=\cN_{2}\dot{W}_{2},\\
\label{eq:cond2}
W_{1}^{2}-\frac{\cN_{1}^{\,2}-1}{12}(2\dot{E}_{1}-E_{1}^{2})-2R
 &=W_{2}^{2}-\frac{\cN_{2}^{\,2}-1}{12}(2\dot{E}_{2}-E_{2}^{2}),
\end{align}
where $R=R_{1}+R_{3}-R_{2}$. The first condition \eqref{eq:cond1}
is easily integrated as
\begin{align}
\label{eq:cond3}
W_{2}=\frac{\cN_{1}}{\cN_{2}}W_{1}+C,
\end{align}
where $C$ is a constant. We note that we can assume
\begin{align}
\label{eq:cond4}
2\dot{E}_{i}-E_{i}^{2}\neq\text{const.}\quad\Longleftrightarrow
 \quad\ddot{E}_{i}-E_{i}\dot{E}_{i}\neq 0;
\end{align}
otherwise, from Eqs.~\eqref{eq:cond2} and \eqref{eq:cond3} we
have $W_{i}=\text{const.}$, which results in a trivial model
$V_{\cN_{1}\!,\,\cN_{2}}^{\pm}=\text{const.}$ Noting further
that $W_{i}$ satisfy Eq.~\eqref{eq:Acon1}, we conclude from
the relation \eqref{eq:cond3} that under the assumption
\eqref{eq:cond4}
\begin{align}
\label{eq:cond5}
2\dot{E}_{1}-E_{1}^{2}=2\dot{E}_{2}-E_{2}^{2}(\neq\text{const.}),
\end{align}
and $C=0$. To investigate the condition \eqref{eq:cond5}, it is
more convenient to recall the fact that the type A Hamiltonians
can be represented as \cite{Ta03a}
\begin{multline}
\label{eq:gHamA}
H^{\pm}=\ee^{-\cW_{\cN_{i}}^{\pm}}\Biggl[-A_{i}(z_{i})
 \frac{\dd^{2}}{\dd z_{i}^{2}}+\biggl(\frac{\cN_{i}-2}{2}A'_{i}(z_{i})
 \pm Q_{i}(z_{i})\biggr)\frac{\dd}{\dd z_{i}}\\
 -\frac{(\cN_{i}-1)(\cN_{i}-2)}{12}A''_{i}(z_{i})\pm\frac{\cN_{i}-1}{2}
 Q'_{i}(z_{i})-R_{i}\Biggr]\ee^{\cW_{\cN_{i}}^{\pm}},
\end{multline}
where the prime denotes derivative with respect to $z_{i}$,
$A_{i}(z_{i})$ and $Q_{i}(z_{i})$ are polynomials of at most
fourth- and second-degree, respectively, and related to $E_{i}(q)$
and $W_{i}(q)$ by
\begin{align}
\label{eq:defAz}
A_{i}(z_{i})&=\frac{1}{2}(\dot{z}_{i})^{2}=a_{4}^{(i)}z_{i}^{4}
 +a_{3}^{(i)}z_{i}^{3}+a_{2}^{(i)}z_{i}^{2}+a_{1}^{(i)}z_{i}
 +a_{0}^{(i)},\\
\label{eq:defA'z}
A'_{i}(z_{i})&=\ddot{z}_{i}=E_{i}\dot{z}_{i},\\
\label{eq:defQz}
Q_{i}(z_{i})&=-W_{i}\dot{z}_{i}=b_{2}^{(i)}z_{i}^{2}+b_{1}^{(i)}z_{i}
 +b_{0}^{(i)}.
\end{align}
The gauge potentials $\cW_{\cN_{i}}^{\pm}$ in Eq.~\eqref{eq:gHamA}
are given by
\begin{align}
\label{eq:gauge}
\cW_{\cN_{i}}^{\pm}=\frac{\cN_{i}-1}{2}\int\dd q\,E_{i}
 \mp\int\dd q\,W_{i}.
\end{align}
{}From the relation \eqref{eq:defA'z}, we obtain
\begin{align}
\label{eq:Hessi}
2\dot{E}_{i}-E_{i}^{2}&=\frac{4H[A_{i}]}{(\dot{z}_{i})^{2}}
 =\frac{1}{(\dot{z}_{i})^{2}}\bigl[4A_{i}A''_{i}-3(A'_{i})^{2}\bigr],\\
\ddot{E}_{i}-E_{i}\dot{E}_{i}&=-\frac{24\,T[A_{i}]}{(\dot{z}_{i})^{3}}
 =\frac{1}{(\dot{z}_{i})^{3}}\bigl[4A_{i}^{2}A_{i}'''-6A_{i}A'_{i}
 A''_{i}+3(A'_{i})^{3}\bigr],
\end{align}
where $H[A_{i}]$ and $T[A_{i}]$ are algebraic covariants called
the Hessian of $A_{i}$ and the Jacobian of $A_{i}$ and $H[A_{i}]$,
respectively \cite{Gu64,Ol99}.\footnote{Here we fix the irrelevant
multiplicative constants for $H[A_{i}]$ and $T[A_{i}]$ after
Refs.~\cite{Ta03a,GKO93}, which are different from those in
Refs.~\cite{Gu64,Ol99}.}
Thus, we see from Eqs.~\eqref{eq:defAz} and \eqref{eq:Hessi} that
the condition \eqref{eq:cond5} is equivalent to $H[A_{1}]/A_{1}=
H[A_{2}]/A_{2}(\neq\text{const.})$. The latter equation is satisfied
if and only if $A_{1}(z_{1})=A_{2}(z_{2})$ and $(\dot{z}_{1})^{2}=
(\dot{z}_{2})^{2}$. Hence, we have\footnote{More precisely, we have
$z_{1}=\pm z_{2}+z_{0}$ where $z_{0}$ is a constant. But we can
always redefine the variables with the aid of the $GL(2,K)$
transformation \eqref{eq:transA} and \eqref{eq:transQ}, under which
the Hamiltonians \eqref{eq:gHamA} are invariant, so that
$z_{1}=z_{2}$.}
\begin{align}
\label{eq:cond6}
z_{1}=z_{2}\equiv z,\qquad A_{1}(z)=A_{2}(z)\equiv A(z).
\end{align}
In this case, Eq.~\eqref{eq:defA'z} implies
\begin{align}
\label{eq:cond7}
E_{1}=E_{2}\equiv E=\frac{\ddot{z}}{\dot{z}}.
\end{align}
{}From this relation together with Eq.~\eqref{eq:cond3} ($C=0$) two
supercharges coincide when $\cN_{1}=\cN_{2}$. Thus, we hereafter
assume $\cN_{1}>\cN_{2}$.
With the use of Eqs.~\eqref{eq:cond3} with $C=0$, \eqref{eq:cond5},
\eqref{eq:defAz}, \eqref{eq:defQz}, and \eqref{eq:Hessi}--%
\eqref{eq:cond7}, the condition \eqref{eq:cond2} and the assumption
\eqref{eq:cond4} are rewritten as
\begin{align}
\label{eq:cond8}
Q_{1}^{2}+\frac{\cN_{2}^{\,2}}{3}H[A]+\frac{4\cN_{2}^{\,2}R}{
 \cN_{1}^{\,2}-\cN_{2}^{\,2}}A=0,\qquad T[A]\neq 0.
\end{align}
We note that the condition \eqref{eq:cond8} is expressed as
algebraically covariant form under the projective transformations
$GL(2,K)$ ($K=\bbR$ or $\bbC$) on $A(z)$ and $Q_{1}(z)$
(cf. Refs.~\cite{Ta03a,GKO93}):
\begin{align}
\label{eq:transA}
A(z)&\mapsto\hat{A}(z)=\Delta^{-2}(\gamma z+\delta)^{4}
 A\biggl(\frac{\alpha z+\beta}{\gamma z+\delta}\biggr),\\
\label{eq:transQ}
Q_{1}(z)&\mapsto\hat{Q}_{1}(z)=\Delta^{-1}(\gamma z+\delta)^{2}
 Q_{1}\biggl(\frac{\alpha z+\beta}{\gamma z+\delta}\biggr),
\end{align}
where $\alpha,\beta,\gamma,\delta\in K$ and $\Delta=\alpha\delta
-\beta\gamma\neq0$. We can easily show that the polynomial $A(z)$
of at most fourth-degree is transformed to one of the five canonical
forms listed in Table~\ref{tb:canon} by the $GL(2,\bbC)$
transformation. Hence, it is sufficient for us to examine each of
the five cases separately.

\begin{table}[ht]
\begin{center}
\arraycolsep = 10pt
\begin{tabular}{llll}
\hline
Case & Canonical Form & $H[A]$ & $T[A]$\\
\hline
I   & $1/2$           & $0$                  & $0$\\
II  & $2z$            & $-3$                 & nonzero\\
III & $2\nu z^{2}$    & $-4\nu^{2}z^{2}$     & $0$\\
IV  & $2\nu(z^{2}-1)$ & $-4\nu^{2}(z^{2}+2)$ & nonzero\\
V   & $2z^{3}-g_{2}z/2-g_{3}/2$
 & $-3z^{3}-3g_{2}z^{2}/2-6g_{3}z-3g_{2}^{2}/16$ & nonzero\\
\hline
\end{tabular}
\end{center}
\vspace{10pt}
\caption{Canonical forms of $A(z)$ and their corresponding $H[A]$ and
 $T[A]$. The parameters $\nu$, $g_{2}$, $g_{3}\in\bbC$ satisfy
 $\nu\neq 0$ and $g_{2}^{3}-27g_{3}^{2}\neq 0$.}
\label{tb:canon}
\end{table}

{}From the last column of Table~\ref{tb:canon}, we see that for
Cases I and III $T[A]=0$ and thus we obtain trivial models. Therefore,
we need to investigate the condition \eqref{eq:cond8} only for Cases
II, IV, and V.

Before proceeding to the case-by-case study, we shall investigate
the solvable sectors $\cV_{\cN_{1}\!,\,\cN_{2}}^{(A)\pm}$ of the type
A $(\cN_{1},\cN_{2})$-fold supersymmetric Hamiltonians
$H_{\cN_{1}\!,\,\cN_{2}}^{(A)\pm}$, namely, the vector spaces preserved
by $H_{\cN_{1}\!,\,\cN_{2}}^{(A)\pm}$. Since type A $\cN_{i}$-fold
supersymmetric Hamiltonian preserves the kernel of type A $\cN$-fold
supercharge, we have
\begin{align}
\label{eq:solV}
\cV_{\cN_{1}\!,\,\cN_{2}}^{(A)\pm}=\cV_{\cN_{1}}^{(A)\pm}\cup
 \cV_{\cN_{2}}^{(A)\pm},
\end{align}
where $\cV_{\cN_{i}}^{(A)\pm}$ are given by
\begin{align}
\cV_{\cN_{i}}^{(A)\pm}=\ker P_{\cN_{i}}^{(A)\pm}
 =\ee^{-\cW_{\cN_{i}}^{\pm}}\bigl\langle 1,z,\dots,z^{\cN_{i}-1}
 \bigr\rangle.
\end{align}
Substituting Eqs.~\eqref{eq:cond3}, \eqref{eq:gauge} and
\eqref{eq:cond7} into the latter equation, we finally obtain
\begin{align}
\cV_{\cN_{1}}^{(A)\pm}&=(\dot{z})^{-\frac{\cN_{1}-1}{2}}\exp\biggl(
 \pm\int\dd q\, W_{1}\biggr)\bigl\langle 1,z,\dots,z^{\cN_{1}-1}
 \bigr\rangle,\\
\cV_{\cN_{2}}^{(A)\pm}&=(\dot{z})^{-\frac{\cN_{2}-1}{2}}
 \exp\biggl(\pm\frac{\cN_{1}}{\cN_{2}}\int\dd q\, W_{1}\biggr)
 \bigl\langle 1,z,\dots,z^{\cN_{2}-1}\bigr\rangle .
\end{align}

\section{Classification of the Models}
\label{sec:class}

In this section, we shall present a detailed classification of
the three non-trivial cases characterized by nonzero $T[A]$ in
Table~\ref{tb:canon}.

\subsection{Case II: $A(z)=2z$, $z(q)=q^{2}$}
\label{ssec:case2}

In this case, the condition \eqref{eq:cond8} is satisfied only if
\begin{align}
b_{2}^{(1)}=b_{1}^{(1)}=0,\quad R=0,\quad b_{0}^{(1)2}=\cN_{2}^{\,2}.
\end{align}
We set $b_{0}^{(1)}=\cN_{2}$ without loss of generality. The functions
$E$ and $W_{1}$ in the supercharge, potentials, and solvable sectors
are given by the followings.

\noindent
\emph{Supercharge:}
\begin{align}
E(q)=\frac{1}{q},\quad W_{1}(q)=-\frac{\cN_{2}}{2q}.
\end{align}
\emph{Potentials:}
\begin{align}
V_{\cN_{1}\!,\,\cN_{2}}^{\pm}(q)=\frac{(\cN_{1}\pm\cN_{2}-1)
 (\cN_{1}\pm\cN_{2}+1)}{8q^{2}}.
\end{align}
\emph{Solvable sectors:}
\begin{align}
\cV_{\cN_{i}}^{(A)\pm}&=q^{-\frac{\cN_{i}\pm\cN_{3-i}-1}{2}}
 \bigl\langle 1,q^{2},\dots,q^{2(\cN_{i}-1)}\bigr\rangle\quad(i=1,2).
\end{align}

\subsection{Case IV: $A(z)=2\nu(z^{2}-1)$, $z(q)=\cosh 2\rnu q$}
\label{ssec:case4}

In this case, the condition \eqref{eq:cond8} is satisfied only if
\begin{equation}
\begin{split}
\label{eq:cond9}
b_{2}^{(1)}=0,\quad b_{1}^{(1)2}-\frac{4\cN_{2}^{\,2}\nu^{2}}{3}
 +\frac{8\cN_{2}^{\,2}R\nu}{\cN_{1}^{\,2}-\cN_{2}^{\,2}}=0,\\
b_{1}^{(1)}b_{0}^{(1)}=0,\quad b_{0}^{(1)2}
 -\frac{8\cN_{2}^{\,2}\nu^{2}}{3}
 -\frac{8\cN_{2}^{\,2}R\nu}{\cN_{1}^{\,2}-\cN_{2}^{\,2}}=0.
\end{split}
\end{equation}
Hence, we shall consider in what follows the two cases separately,
namely, $b_{0}^{(1)}=0$ first and $b_{1}^{(1)}=0$ next. When
$b_{0}^{(1)}=0$, the condition \eqref{eq:cond9} is equivalent to
\begin{align}
b_{0}^{(1)}=0,\quad b_{1}^{(1)2}=4\cN_{2}^{\,2}\nu^{2},\quad
 R=-\frac{\cN_{1}^{\,2}-\cN_{2}^{\,2}}{3}\nu.
\end{align}
We set $b_{1}^{(1)}=2\cN_{2}\nu$ without loss of generality.
The functions $E$ and $W_{1}$ in the supercharge, potentials,
and solvable sectors are given by the followings.

\noindent
\emph{Supercharge:}
\begin{align}
E(q)=\frac{2\rnu\cosh 2\rnu q}{\sinh 2\rnu q},\quad
 W_{1}(q)=-\frac{\cN_{2}\rnu\cosh 2\rnu q}{\sinh 2\rnu q}.
\end{align}
\emph{Potentials:}
\begin{align}
V_{\cN_{1}\!,\,\cN_{2}}^{\pm}=\frac{(\cN_{1}\pm\cN_{2}-1)
 (\cN_{1}\pm\cN_{2}+1)\nu}{2\sinh^{2}2\rnu q}+\frac{3\cN_{1}^{\,2}
 +\cN_{2}^{\,2}-1}{6}\nu.
\end{align}
\emph{Solvable sectors:}
\begin{align}
\cV_{\cN_{i}}^{(A)\pm}=&\,(\sinh 2\rnu q)^{
 -\frac{\cN_{i}\pm\cN_{3-i}-1}{2}}\notag\\
&\,\times\bigl\langle 1,\cosh 2\rnu q,\dots,(\cosh 2\rnu q)^{\cN_{i}-1}
 \bigr\rangle\quad(i=1,2).
\end{align}

Next, when $b_{1}^{(0)}=0$, the condition \eqref{eq:cond9} is
equivalent to
\begin{align}
b_{1}^{(1)}=0,\quad b_{0}^{(1)2}=4\cN_{2}^{\,2}\nu^{2},\quad
 R=\frac{\cN_{1}^{\,2}-\cN_{2}^{\,2}}{6}\nu.
\end{align}
We set $b_{0}^{(1)}=2\cN_{2}\nu$ without loss of generality.
The functions $E$ and $W_{1}$ in the supercharge, potentials,
and solvable sectors are given by the followings.

\emph{Supercharge:}
\begin{align}
E(q)=\frac{2\rnu\cosh 2\rnu q}{\sinh 2\rnu q},\quad
 W_{1}(q)=-\frac{\cN_{2}\rnu}{\sinh 2\rnu q}.
\end{align}
\emph{Potentials:}
\begin{align}
V_{\cN_{1}\!,\,\cN_{2}}^{\pm}=\frac{(\cN_{1}\mp\cN_{2}-1)(\cN_{1}
 \mp\cN_{2}+1)\nu}{2\sinh^{2}2\rnu q}\pm\frac{\cN_{1}\cN_{2}
 \nu}{2\sinh^{2}\rnu q}+\frac{\cN_{2}^{\,2}-1}{6}\nu.
\end{align}
\emph{Solvable sectors:}
\begin{align}
\cV_{\cN_{i}}^{(A)\pm}=&\,(\sinh 2\rnu q)^{-\frac{\cN_{i}-1}{2}}
 (\tanh\rnu q)^{\mp\frac{\cN_{3-i}}{2}}\notag\\
&\,\times\bigl\langle 1,\cosh 2\rnu q,\dots,
 (\cosh 2\rnu q)^{\cN_{i}-1}\bigr\rangle\quad(i=1,2).
\end{align}

\subsection{Case V: $A(z)=2z^{3}-g_{2}z/2-g_{3}/2$, $z(q)=\wp(q)$}
\label{ssec:case5}

In this case, the condition \eqref{eq:cond8} is satisfied only if
\begin{equation}
\begin{split}
b_{2}^{(1)2}=\cN_{2}^{\,2},\quad 2b_{2}^{(1)}b_{1}^{(1)}
 +\frac{8\cN_{2}^{\,2}R}{\cN_{1}^{\,2}-\cN_{2}^{\,2}}=0,\quad
 2b_{2}^{(1)}b_{0}^{(1)}+b_{1}^{(1)2}-\frac{\cN_{2}^{\,2}g_{2}}{2}=0,\\
2b_{1}^{(1)}b_{0}^{(1)}-2\cN_{2}^{\,2}g_{3}-\frac{2\cN_{2}^{\,2}R
 g_{2}}{\cN_{1}^{\,2}-\cN_{2}^{\,2}}=0,\quad b_{0}^{(1)2}
 -\frac{\cN_{2}^{\,2}g_{2}^{2}}{16}-\frac{2\cN_{2}^{\,2}R g_{3}}{
 \cN_{1}^{\,2}-\cN_{2}^{\,2}}=0.
\end{split}
\end{equation}
We set $b_{2}^{(1)}=\cN_{2}$ without loss of generality. Then,
the latter set of conditions is equivalent to
\begin{gather}
\label{eq:cond10}
b_{2}^{(1)}=\cN_{2},\quad b_{1}^{(1)}=-\frac{4\cN_{2}}{\cN_{1}^{\,2}
 -\cN_{2}^{\,2}}R,\quad b_{0}^{(1)}=-\frac{\cN_{2}}{4}g_{2}
 -\frac{\cN_{2}(\cN_{1}^{\,2}-\cN_{2}^{\,2})}{4R}g_{3},\\
\label{eq:cond11}
\frac{32R^{3}}{(\cN_{1}^{\,2}-\cN_{2}^{\,2})^{3}}-\frac{2R}{\cN_{1}^{
 \,2}-\cN_{2}^{\,2}}g_{2}-g_{3}=0.
\end{gather}
If we introduce the values of the Weierstrass function $\wp(q)$ at
the half of the fundamental periods $2\omega_{l}$
\begin{align}
e_{l}=\wp(\omega_{l})\quad (l=1,2,3),
\end{align}
which all satisfy the algebraic equation of third-degree $4e_{l}^{3}
-e_{l}g_{2}-g_{3}=0$, a solution of Eq.~\eqref{eq:cond11} is
represented as
\begin{align}
R=\frac{\cN_{1}^{\,2}-\cN_{2}^{\,2}}{2}e_{k},
\end{align}
where $k=1$, $2$, or $3$. Substituting the latter expression for $R$
into Eq.~\eqref{eq:cond10}, we obtain
\begin{align}
b_{1}^{(1)}=-2\cN_{2}e_{k},\quad b_{0}^{(1)}=-\cN_{2}
 (H_{k}^{2}-e_{k}^{2}),
\end{align}
where $H_{l}^{2}$ is defined by
\begin{align}
H_{l}^{2}=3e_{l}^{2}-\frac{g_{2}}{4}=(e_{l}-e_{m})(e_{l}-e_{n})
 \quad (l=1,2,3;\, l\neq m\neq n\neq l).
\end{align}
Finally, The functions $E$ and $W_{1}$ in the supercharge, potentials,
and solvable sectors are given by the followings.

\noindent
\emph{Supercharge:}
\begin{align}
E(q)=\frac{12\wp(q)^{2}-g_{2}}{2\wp'(q)},\quad W_{1}(q)=-\cN_{2}
 \frac{\wp(q)^{2}-2e_{k}\wp(q)-H_{k}^{2}+e_{k}^{2}}{\wp'(q)}.
\end{align}
\emph{Potentials:}
\begin{align}
V_{\cN_{1},\cN_{2}}^{\pm}=&\,\frac{(\cN_{1}\mp\cN_{2}-1)(\cN_{1}\mp
 \cN_{2}+1)}{8}\biggl(\wp(q)+\sum_{l=1}^{3}\frac{H_{l}^{4}-18e_{l}^{2}
 H_{l}^{2}+36e_{l}^{4}}{H_{l}^{2}(\wp(q)-e_{l})}\biggr)\notag\\
&\,\mp\frac{\cN_{1}\cN_{2}}{2}\sum_{l=1}^{3}\frac{e_{k}^{2}H_{l}^{2}
 +(e_{k}+2e_{l})e_{l}(5H_{l}^{2}-12e_{l}^{2})}{H_{l}^{2}(\wp(q)-e_{l})}
 -\frac{\cN_{1}(\cN_{1}\pm\cN_{2})}{2}e_{k}.
\end{align}
\emph{Solvable sectors:}
\begin{align}
\cV_{\cN_{i}}^{(A)\pm}=&\,\prod_{l=1}^{3}\bigl(\wp(q)-e_{l}\bigr)^{
 -\frac{\cN_{i}-1}{4}\mp\cN_{3-i}\frac{(e_{k}+e_{l})^{2}-H_{k}^{2}}{
 4H_{l}^{2}}}\notag\\
&\,\times\bigl\langle 1,\wp(q),\dots,\wp(q)^{\cN_{i}-1}
 \bigr\rangle\quad(i=1,2).
\end{align}

\section{Concluding Remarks}
\label{sec:concl}

In this article, we have solved the condition for simultaneous
$\cN$-fold supersymmetry with respect to two type A $\cN$-fold
supercharges and classified all the possible models. It turns
out that there are essentially four inequivalent models possessing
the property, the first is conformal, the second is Rosen--Morse
type, the third is another hyperbolic ($\nu>0$) or trigonometric
($\nu<0$) type, and the last is elliptic. However, it does not
mean that we have exhausted the investigation for
$(\cN_{1},\cN_{2})$-fold supersymmetric models. Indeed, it is
interesting to study possibility for another family of simultaneous
$\cN$-fold supersymmetry, namely, with respect to two type B, two
type C, or two different types of $\cN$-fold supercharges.

One may be curious about the relation between type A
$(\cN_{1},\cN_{2})$-fold supersymmetry investigated here and type C
$\cN$-fold supersymmetry with $\cN=\cN_{1}+\cN_{2}$ in
Ref.~\cite{GT05} since they are similar in the sense that both of
them preserve two type A monomial spaces of dimension $\cN_{1}$ and
$\cN_{2}$. In this respect, we would like to stress at least three
significant differences between them. First, a pair of
Hamiltonians of the former is related by $\cN_{1}$th- and
$\cN_{2}$th-order differential operators while that of the latter is
by a differential operator of order $\cN_{1}+\cN_{2}$. Second,
the gauge potentials which connect the spaces acted by the
Hamiltonians and the monomial spaces are different between the former
and the latter systems. Third, solvable sectors of type C
$\cN$-fold supersymmetry always decomposes as a direct sum of two
spaces while those of type A $(\cN_{1},\cN_{2})$-fold supersymmetry
may not, that is, the two spaces which together constitute it
through Eq.~\eqref{eq:solV} can have nontrivial intersection,
$\cV_{\cN_{1}}^{(A)\pm}\cap\cV_{\cN_{2}}^{(A)\pm}\neq\emptyset$.

In this article, we have not touched upon underlying mathematical
details such as those investigated in Ref.~\cite{AS03} since our
present motivation rather comes from physical applications. Thus,
it is interesting to see, for instance, what kind of supercharges
we will obtain for our case after the process of the
\emph{optimization of supercharges} in Ref.~\cite{AS03}.

Finally, we note that we have not restricted the potentials to be real
so that we could classify the models into less number of inequivalent
classes. Hence, some of the models presented here may not describe
a realistic physical system, which is one of the reason why we have
avoided to discuss normalizability of the solvable sectors. In this
respect, we have found that a certain subclass of the present models
would actually fit in describing a physical system and would lead
to dynamical (non-)breaking of $\cN$-fold supersymmetry, which we
would like to report in a subsequent publication \cite{HT05b}.

\begin{ack}
This work was partially supported by the National Science Council
of the Republic of China under the Grant No. NSC-93-2112-M-032-009.
\end{ack}

\bibliography{refsels}
\bibliographystyle{npb}

\end{document}